\documentclass[%
 reprint,
superscriptaddress,
showpacs,
showkeys,
amsmath,amssymb,
aps,
pra,
]{revtex4-1}
\usepackage{graphicx,amssymb,amsmath,epsfig}
\usepackage{color}

\begin{document}

\title{Structural and Electronic Properties of Amorphous Silicon and Germanium Monolayers and Nanotubes: A DFT Investigation}
\author{Raphael M. Tromer}
\affiliation{State University of Campinas, Applied Physics Department, Campinas, São Paulo, 13083-970, Brazil.}

\author{Marcelo L. Pereira J\'unior}
\affiliation{University of Brasília, Faculty of Technology, Department of Electrical Engineering, Brasília, Federal District, 70919-970, Brazil.}

\author{Luiz. A. Ribeiro Júnior}
\affiliation{University of Bras{\'{i}}lia, Institute of Physics, Brasília, Federal District, 70919-970, Brazil}
\affiliation{Computational Materials Laboratory, LCCMat, Institute of Physics, University of Brasília, Brasília, Federal District, 70919-970, Brazil.}

\author{Douglas S. Galv\~ao}
\affiliation{State University of Campinas, Applied Physics Department, Campinas, São Paulo, 13083-970, Brazil.}

\date{\today}

\begin{abstract}
   A recent breakthrough has been achieved by synthesizing monolayer amorphous carbon (MAC), which introduces a material with unique optoelectronic properties. Here, we used ab initio (DFT) molecular dynamics simulations to study silicon and germanium MAC analogs. Typical unit cells contain more than 600 atoms. We also considered their corresponding nanotube structures. The cohesion energy values for MASi and MAGe range from -8.41 to -7.49 eV/atom and follow the energy ordering of silicene and germanene. Their electronic behavior varies from metallic to small band gap semiconductors. Since silicene, germanene, and MAC have already been experimentally realized, the corresponding MAC-like versions we propose are within our present synthetic capabilities. 
\end{abstract}

\pacs{}

\keywords{Monolayer Amorphous Carbon (MAC), Two-dimensional Materials, Density Functional Theory (DFT), Electronic Properties.}

\maketitle

\section{Introduction}
Since the experimental realization of graphene in 2004, the ongoing search for new two-dimensional (2D) materials has been receiving increased attention and has already produced interesting results \cite{Lemme2022, Xie2022, Novoselov2016, Tromer2020, Tromer2021, Tromer2023, Zeng2018}. Generally, ideal material candidates for prospective applications should be easily synthesizable \cite{Kumbhakar2023, Khan2020, Huang2022} and preferably cost-effective \cite{Buapan2021, Zhao2020}.

The growing interest in prototype graphene-based devices has produced several theoretical and experimental investigations \cite{Li2020, Wang2017}. The aim is to obtain novel structures and advance the fabrication of more efficient materials \cite{shanmugam2022review,momeni2020multiscale,kumbhakar2023prospective}. A recent milestone along these lines was the successful synthesis of monolayer amorphous carbon (MAC) by Toh and collaborators \cite{Toh2020SynthesisAP}. MAC is an amorphous material with sp$^2$ and sp$^3$ carbon atoms organized in a 2D lattice. MAC exhibits a wide distribution of bond lengths and angle values, including rings of five to eight atoms. It is worth highlighting that, although short-range order is present in MAC, the interatomic distances and inter-bonding angles deviate from those observed in pristine graphene lattices.

The MAC stable structure was synthesized through a straightforward method involving laser-assisted chemical vapor deposition, standing large structural deformations at high load strain values without fracture or crack propagation \cite{Toh2020SynthesisAP}. Since its synthesis, MAC has been comprehensively investigated by various research groups, leading to a better understanding of its properties and potential applications \cite{garzon2022optoelectronic,tian2023disorder,gastellu2022electronic,xie2021roughening,kilgour2020generating,xie2023toughening,zhang2022structure,zhang2022thermal,Felix2020,PereiraJunior2021,Tromer_mac_2021}. Theoretical investigations have probed the MAC mechanical properties. MAc's yield strength is relatively high compared to other two-dimensional carbon allotropes \cite{Felix2020, PereiraJunior2021, Tromer_mac_2021}. It should be stressed that although other 2D carbon amorphous structures have been proposed in the literature \cite{kotakoski2011point,bhattarai2018amorphous,doyle1995vibrational}, MAC exhibits a distinctive ring distribution, establishing its uniqueness among 2D carbon structures \cite{an2023ultrathin}. 

Among these studies, a Monte Carlo algorithm was employed to distribute B and N atoms into an initial MAC sheet, yielding a distinct structure named 2D monolayer amorphous boron nitride (MABN). They monitored the MABN dynamics as a function of temperature and intrinsic amorphous characteristics \cite{zhang2022structure}. The results revealed that MABN exhibits insulating behavior and bond lengths of approximately 1.54 \r{A}, attributed to N-N and B-B bonds. The same research group performed reverse nonequilibrium molecular dynamics (RNEMD) simulations, demonstrating that the thermal conductivity at 300 K for MAC and MABN are 13.3 and 7.2 $\mathrm{W}\cdot\mathrm{m}^{-1}\cdot\mathrm{K}^{-1}$, respectively \cite{zhang2022thermal}. However, MAC-based sheets and nanotubes formed with other chemical species from group 14 have not been reported in the literature, and the present work aims to fill this gap.

In this study, we propose monolayers and nanotubes MAC-based analogs composed of silicon and germanium atoms. These structures were obtained using the reported MAC structural model \cite{Toh2020SynthesisAP} as a starting point, replacing the carbon atoms with silicon (MASi) and germanium (MAGe). Their structural and electronic properties were investigated using DFT methods. Their electronic band structures and frontier orbital patterns are contrasted against those obtained for the MAC. Additionally, we explore the differences between these properties when the systems are rolled up to form nanotubes.

\section{Methods}

\begin{figure}[]
    \centering
    \includegraphics[scale=0.23]{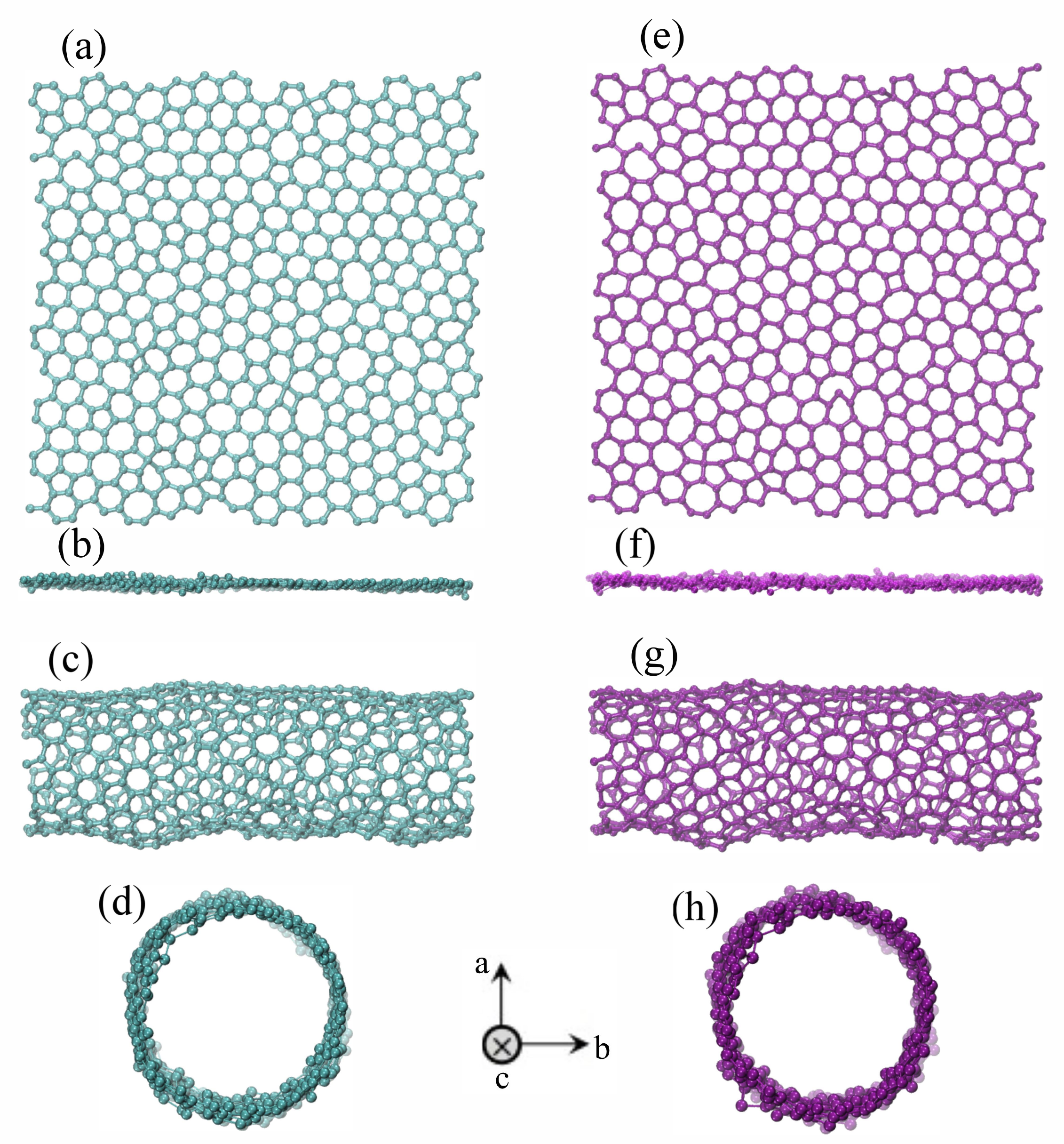}
    \caption{MASi and MAGe optimized unit cell structural models. The left panels (a-d) depict the MASi system, with (a) top and (b) side views of nanosheets, followed by (c) longitudinal and (d) frontal views of tubes. The right panels (e-h) illustrate the corresponding MAGe systems. The axial reference system indicates the crystallographic directions.}
    \label{fig:structures}
\end{figure}

For creating the MASi and MAGe structural models, all carbon atoms in the previously discussed MAC structure were replaced with silicon and germanium atoms, respectively. Subsequently, an optimization methodology was used, involving the simultaneous relaxation of both atoms and lattice vectors along the $x$ and $y$ directions (and $z$ for tubes), considering the systems' periodic boundary conditions. It is worth mentioning that DFT and molecular dynamics simulations of silicene and germanene (the silicon and germanium equivalent of graphene) show significant structural buckling \cite{YanVoon2014}. The atoms deviate from the plane, unlike graphene. These have been experimentally validated \cite{oughaddou2015silicene,hastuti2020first}. 

To investigate the electronic properties and structural stability of the MASi/MAGe nanostructures, we performed first-principles calculations (DFT), as implemented in the SIESTA code \cite{Soler2002,artacho1999linear,artacho2008siesta}. The exchange–correlation term was modeled using the generalized gradient approximation (GGA-PBE) \cite{perdew1996generalized}. The wave functions were represented as a linear combination of atomic orbitals, employing a summation of a double zeta basis set along with a polarization function (DZP).

The interaction between valence electrons and atomic ions was modeled using the norm-conserving Troullier–Martins pseudopotential, adopting a Bylander-factorized form \cite{troullier1991efficient}. In the reciprocal space, we employed a Monkhorst–Pack scheme at the $\Gamma$ k-point, given the large size of the cell corresponding to 610 atoms, for electronic analyses, such as band structure calculations, a mesh of $4\times 4\times 1$ ($1\times 1\times 4$) was considered for nanosheets (nanotubes). Calculations were performed with an energy cut-off of $200$ Ry.

Convergence in self-consistent calculation cycles was established when the maximum difference between density matrix elements was less than $10^{-4}$ eV. During the structural optimization, atoms and lattice parameters could vary until the force acting on each atom reached a magnitude less than $0.05$ eV/\r{A}.

To investigate potential changes at room temperature and, most importantly, to reliably describe the buckling states, ab initio molecular dynamics (AIMD) simulations were performed at $T=300$ K (see Figure \ref{fig:structures}). A time step of 1.0 fs was employed with an NVT ensemble for a total simulation duration of 500 steps. A Nosé–Hoover thermostat was used to control the temperature of the AIMD simulations. It is worth stressing that the final structures derived from AIMD simulations were used as input for the cohesive energy, band structure, and frontier orbitals calculations.

To estimate the cohesive energy, the following relation was used:
\begin{equation}
E_\textrm{coh}=\frac{E_\textrm{MASi/MAGe}-N_\mathrm{Si/Ge}\cdot E_\mathrm{Si/Ge}}{N_\mathrm{Si/Ge}}.
\label{eq:cohesive}
\end{equation}
Here, $E_\mathrm{MASi/MAGe}$ represents the MASi/MAGe total energy, $E_\mathrm{Si/Ge}$ is the energy of an isolated Si/Ge atom, and $N_\mathrm{Si/Ge}$ denotes the total number of atoms.

\section{Results}

In Figure \ref{fig:structures}, we present the optimized structures obtained from the AIMD simulations. MASi-based structures are shown on the left (panels (a) to (d)), while the MAGe-ones are shown on the right (panels (e) to (h)). The sheets exhibit the expected buckling, where the atoms are slightly above or below the two-dimensional plane. This lattice arrangement differs from the more pronounced buckling observed in silicene or germanene \cite{oughaddou2015silicene,hastuti2020first}. The presence of lattice structural disorder significantly decreases the buckling effect. These results agree with previous studies in which the impact of carbon doping, defects, and network size on silicene structures was explored \cite{Pablo-Pedro2020}.

The results in Figure \ref{fig:structures} suggest that MA(Si/Ge) systems tend to be less sensitive to changes in out-of-plane atomic displacements by decreasing the intrinsic buckling. Although minimal, buckling effects are slightly more pronounced in germanium-based systems \cite{nijamudheen2015electronic,jose2012understanding}. This trend aligns with the fact that germanium, having a larger atomic radius, would exhibit more pronounced out-of-plane atomic displacements \cite{Balendhran2015}. More importantly, this also reflects that silicon and germanium exhibit the so-called pseudo-Jahn-Teller effect \cite{Eric}, which favors the sp$^3$ hybridization, while carbon favors the sp$^2$ one.

In the case of nanotubes (refer to Figures \ref{fig:structures}(c-d) and (g-h)), the buckling effect is slightly more pronounced than for the nanosheet cases. This result suggests that curvature effects play a crucial role in inducing the buckling states in these systems. Hence, there is a competition between disorder and curvature, where one tends to decrease, and the other tends to increase the structural buckling. The buckling level can be correlated to changes in the electronic changes induced by external disorder factors \cite{Pablo-Pedro2020,Eric}. In this way, for MA(Si/Ge) systems, the lattice curvature has a more pronounced influence on the electronic structure regarding the disorder than for the carbon structures.

\begin{table}[]
    \centering
    \caption{Structural parameters for MASi and MAGe nanosheets and nanotubes.}
    \label{tab:parmaeters}
    \begin{tabular}{|c|c|c|c|c|}
    \hline
        Structures & MASi & MAGe & MASi-NT & MAGe-NT  \\
         \hline
         E$_\textrm{coh}$ (eV/atom)&-8.40&-7.49&-8.41&-7.50\\
         \hline
         $a$ (\AA)&61.78&63.40&30.00&30.00\\
         \hline
         $b$ (\AA)&61.83&63.44&30.00&30.00\\
         \hline
         $c$ (\AA)&30.00&30.00&61.17&62.65\\
         \hline
         $\alpha~(^\circ)$&87.98&88.93&90.33&91.11\\
         \hline
         $\beta~(^\circ)$&89.98&89.87&89.84&89.18\\
         \hline
         $\gamma~(^\circ)$&89.96&90.05&91.92&92.15\\
         \hline
         $\overline{d}_\textrm{bond}$&2.21&2.31&2.21&2.32\\
         \hline
    \end{tabular}
\end{table}

In addition to the discussion on buckling, Table \ref{tab:parmaeters} presents other structural characteristics of interest for the investigated systems. The lattice parameters ($a$, $b$, $c$, $\alpha$, $\beta$, and $\gamma$) for germanium are more significant than those for silicon in both layers and tubes, as expected due to the greater atomic radius of germanium compared to silicon. The average bond lengths ($\overline{d}_\textrm{bond}$) for both layers and tubes are practically the same: $2.21$ \r{A} for silicon and $2.31$ \r{A} for germanium, indicating that although the curvature affects the buckling, it does not significantly affect the bond lengths. These values are in agreement with those obtained for silicene and germanene, which are reported as $2.25$ \r{A} and $2.38$ \r{A}, respectively \cite{Takahashi2021,Dimoulas2015}.

In Table \ref{tab:parmaeters}, we summarize the structural features and the cohesive energy values obtained for each system using Equation \ref{eq:cohesive}, which correlates the system's structural stability. The Table shows that the cohesive energies for the silicon layer and tube are smaller than the values for germanium analogs, with differences of almost $1.0$ eV. This trend is consistent with the relative stability trend observed for silicene and germanene \cite{Yang2022}.

\begin{figure}[]
    \centering
    \includegraphics[width=\linewidth]{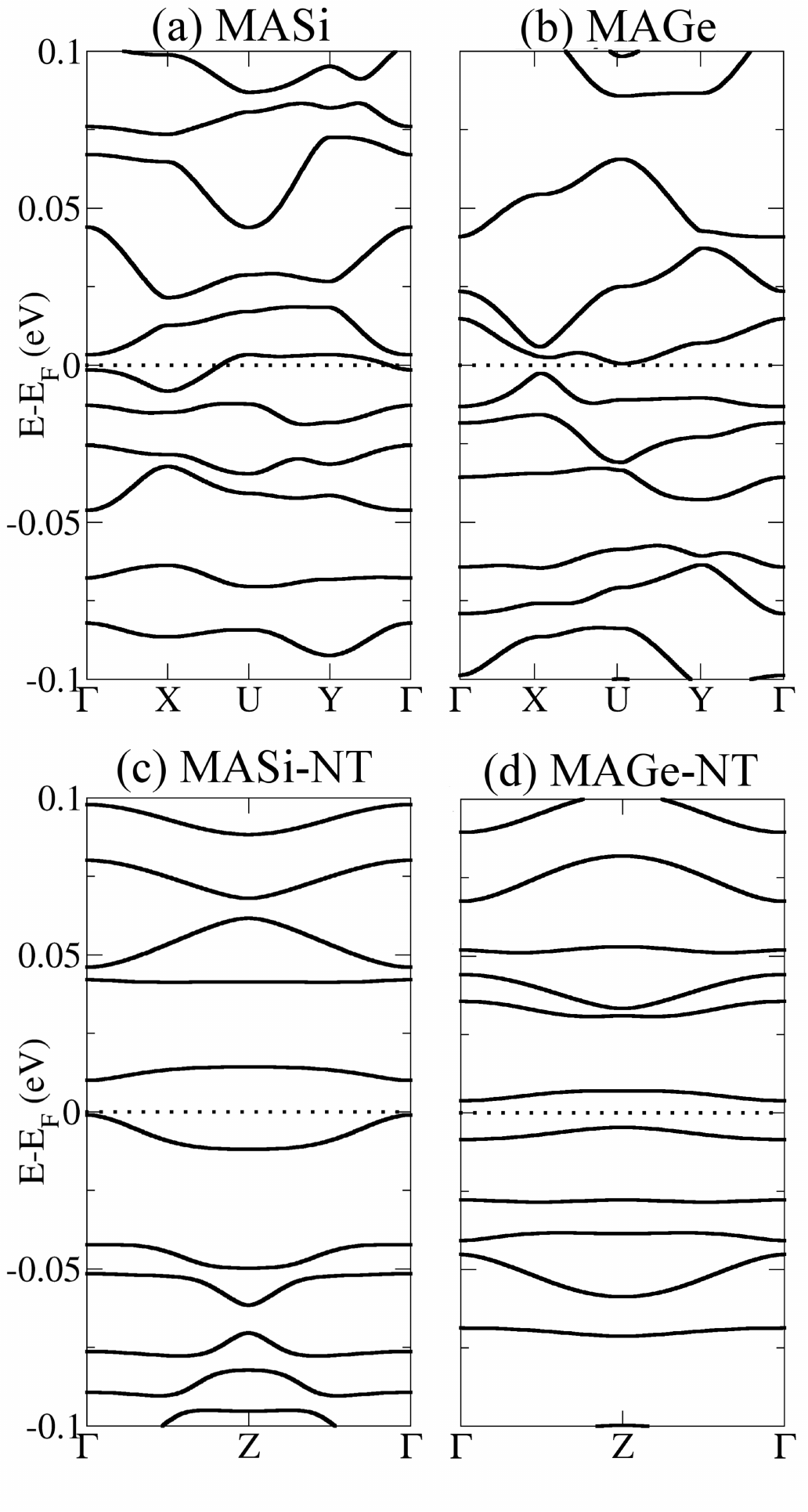}
    \caption{Electronic band structure calculations for the investigated structures: (a) MASi, (b) MAGe, (c) MASi-NT, and (d) MAGe-NT.}
    \label{fig:bands}
\end{figure}

In Figure \ref{fig:bands}, we present the electronic band structure for the investigated systems. In the case of MASi, valence band crossings occur at the Fermi level, indicating metallic characteristics (see Figure \ref{fig:bands}(a)). These crossings are observed between the symmetry points of the square lattice, between X-U and Y-$\Gamma$. In contrast, all other structures are low-band-gap semiconductors with small band gaps. The calculated electronic band gap values are 3.2 (indirect, Figure \ref{fig:bands}(b)), 11.1 (direct, Figure \ref{fig:bands}(c)), and 8.4 meV (indirect, Figure \ref{fig:bands} (d)) for MAGe, MASi-NT and MAGe-NT, respectively.

It is well-known that GGA/PBE calculations underestimate electronic band gaps \cite{10.1063/1.358463}. These factors, when combined, may lead to a transition from a low-band gap semiconductor to a material with semi-metallic or even metallic characteristics. We can also observe from Figure \ref{fig:bands} that nanotubes exhibit a slightly larger band gap value than the corresponding sheets. This trend is consistent with previous investigations on silicene, where nanotube chirality was also found to play a crucial role in determining the band gap value due to variations in the overlap of molecular orbitals \cite{Liu2021}. Also, as expected, the nanotube topology induces electronic confinement, which also contributes to increasing the bandgap values.

\begin{figure}[]
    \centering
    \includegraphics[width=\linewidth]{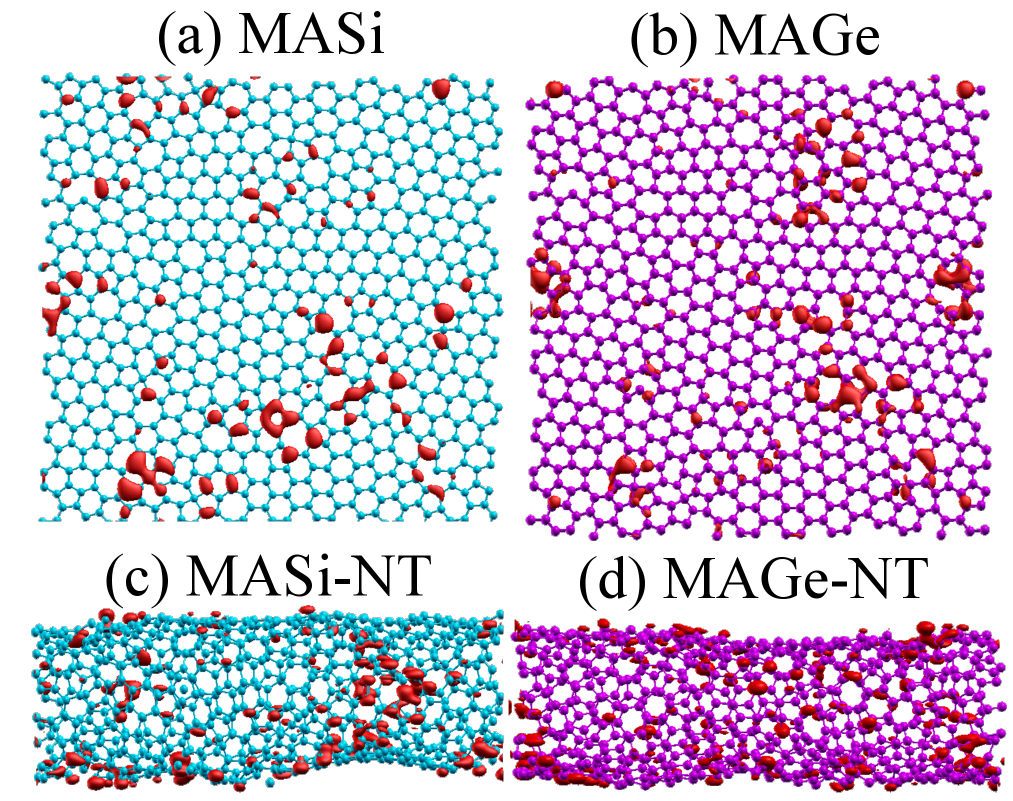}
    \caption{Lowest Unoccupied Crystal Orbital (LUCO) for MASi (a), MASi-NT (b), MAGe (c), and MAGe-NT (d).}
    \label{fig:luco}
\end{figure}

In Figure \ref{fig:luco}, we present the squared modulus of the wave function corresponding to the lowest unoccupied crystalline orbitals (LUCO), for both MA(Si/Ge) for the sheets and tubes. We can see that LUCO in all systems has a significant localization on the nonhexagonal rings separated by crystalline regions. A subtle difference in localization between MASi and MAGe is observed due to their distinct disorder patterns. These disorders are attributed to the reduced number of hexagonal rings. It is worth noting that the highest occupied crystalline orbitals (HOCO) localization trends for MA(Si/Ge) show the same behavior as the LUCO. These localization trends among LUCO and HOCO are analogous to what happens in the MAC case \cite{Tromer_mac_2021}.

\section{Conclusions}

In summary, we carried out DFT simulations to investigate the structural and electronic properties of the silicon and germanium versions of the experimentally realized monolayer amorphous carbon. Our findings reveal that these materials in the nanosheets and nanotubes form can present metallic and narrow semiconducting electronic band gaps. Cohesive energy values for the Si and Ge-based structures follow the same ordering for silicene and germanene. Frontier orbitals analysis revealed similar electronic localization for LUCO and HOCO. These localizations are related to the arrangement of nonhexagonal rings separated by crystalline regions. 

Since silicene, germanene, and MAC have already been experimentally realized, their corresponding MAC-like versions are within our present synthetic capabilities. We hope the present work will stimulate further studies on this new class of nanostructures.  

\section*{Acknowledgements}

This work received partial support from Brazilian agencies CAPES, CNPq, and FAPDF. We thank the Center for Computing in Engineering and Sciences at Unicamp for financial support through the FAPESP/CEPID Grants \#2013/08293-7 and \#2018/11352-7. M.L.P.J acknowledges the financial support from FAP-DF grant 00193-00001807/2023-16. L.A.R.J. acknowledges CAPES for partially financing this study—Finance Code 88887.691997/2022-00. CENAPAD-SP (Centro Nacional de Alto Desenpenho em São Paulo – Universidade Estadual de Campinas—UNICAMP) provided computational support for M.L.P.J and L.A.R.J (proj960, and proj634). L.A.R.J thanks the financial support from the Brazilian Research Council FAP-DF grant 00193.00001808/2022-71, FAPDF-PRONEM grant 00193.00001247/2021-20, 00193-00001857/2023-95, 00193-00001782/2023-42, and CNPq grants 350176/2022-1. L.A.R.J and M.L.P.J also thanks Núcleo de Computação de Alto Desempenho (NACAD) for computational facilities through the Lobo Carneiro Supercomputer. 

\bibliography{References}

\end{document}